\documentclass[twocolumn,
superscriptaddress,
amsmath,amssymb,
]{revtex4-2}
\usepackage{xcolor}
\usepackage[normalem]{ulem}
\usepackage{cancel}
\usepackage{graphicx}
\usepackage{enumerate}
\usepackage{tikz}
\usetikzlibrary{arrows.meta,positioning,decorations.pathreplacing,calc}
\usepackage{soul}
\definecolor{del}{HTML}{00008A}
\definecolor{add}{HTML}{8A0000}

\usepackage[colorlinks=true,allcolors=blue]{hyperref}
\definecolor{ReadBackground}{HTML}{EFEFEF}

\begin{document}
\title{Architectural scaling tradeoffs in modular 3D bosonic quantum processors}
\author{Shaojiang Zhu}
\email{szhu26@fnal.gov}
\affiliation{Superconducting Quantum Materials and Systems Center, Fermi National Accelerator Laboratory, Batavia, IL 60510, USA}%
\author{Ugur Alyanak}
\affiliation{Superconducting Quantum Materials and Systems Center, Fermi National Accelerator Laboratory, Batavia, IL 60510, USA}
\affiliation{Department of Physics, University of Chicago, Chicago, IL 60637, USA}
\author{Tanay Roy}
\affiliation{Superconducting Quantum Materials and Systems Center, Fermi National Accelerator Laboratory, Batavia, IL 60510, USA}
\author{Alessandro Reineri}
\affiliation{Superconducting Quantum Materials and Systems Center, Fermi National Accelerator Laboratory, Batavia, IL 60510, USA}
\author{Andy C. Y. Li}
\affiliation{Superconducting Quantum Materials and Systems Center, Fermi National Accelerator Laboratory, Batavia, IL 60510, USA}
\author{Taeyoon Kim}
\affiliation{Superconducting Quantum Materials and Systems Center, Fermi National Accelerator Laboratory, Batavia, IL 60510, USA}
\author{Srivatsan Chakram}
\affiliation{Department of Physics and Astronomy, Rutgers University, Piscataway, New Jersey, 08854, USA.}
\author{Akshay Murthy}
\affiliation{Superconducting Quantum Materials and Systems Center, Fermi National Accelerator Laboratory, Batavia, IL 60510, USA}
\author{Anna Grassellino}
\affiliation{Superconducting Quantum Materials and Systems Center, Fermi National Accelerator Laboratory, Batavia, IL 60510, USA}
\author{Alexander Romanenko}
\affiliation{Superconducting Quantum Materials and Systems Center, Fermi National Accelerator Laboratory, Batavia, IL 60510, USA}

\begin{abstract}
We propose a modular three-dimensional bosonic quantum processor built from repeatable coupled-cavity modules linked by configurable interconnect networks. 
Using hardware-motivated graph-theoretic measures, we compare nearest-neighbor, hub-based, and hybrid architectures in terms of interconnect count, communication distance, resource concentration, and implementation complexity. 
Rather than identifying a universally optimal topology, our analysis shows how these architectures redistribute the costs of scaling, including wiring and port requirements, nonlocal communication distance, exposure to shared resources, routing bottlenecks, and scheduling overhead. 
Case studies of a \(3\times3\) processor and a larger hierarchical architecture further distinguish finite-size performance from asymptotic scaling. 
The resulting framework provides a systematic basis for evaluating modular three-dimensional bosonic processors and for identifying the device-level parameters required for quantitative hardware design.
\end{abstract}

\maketitle
\section{Introduction}\label{sec:introduction}
Superconducting quantum circuits are a promising platform for quantum information processing because they combine strong nonlinearity, flexible circuit design, and compatibility with microwave control and readout~\cite{devoret2013superconducting}. 
Cavity-based and bosonic approaches are particularly attractive because long-lived electromagnetic modes can serve as robust quantum memories and computational resources~\cite{devoret2007circuit, blais2021circuit, gambetta2017building}. 
Scaling these processors beyond small and intermediate system sizes, however, remains challenging. 
In monolithic architectures, increasing processor size places growing demands on wiring, packaging, mode management, crosstalk suppression, calibration, and fabrication yield~\cite{bardin2021microwaves, stanley2025review, krinner2019engineering, krasnok2024superconducting, ketterer2023characterizing}. 
These constraints can become limiting even when individual components perform well.

Three-dimensional (3D) integration offers important advantages in this context. 
Compared with planar implementations, 3D cavity-based structures provide longer coherence, improved electromagnetic isolation, and reduced participation in selected loss channels~\cite{paik2011observation, rosenberg20173d, romanenko2020three, oriani2025niobium}. 
These properties make them a natural platform for bosonic quantum processing. 
Nevertheless, 3D integration alone does not resolve the scaling challenges of large monolithic systems. 
As the processor size grows, global wiring access, interconnect complexity, and package-level mode control can again become bottlenecks~\cite{pietikainen2024strategies}. 
Moreover, demonstrations of bosonic error correction, remote state transfer, and operations between distinct cavity modes show that scalable processing requires communication and control resources beyond individual computational modes~\cite{ofek2016extending, axline2018demand, burkhart2021error, rosenblum2018cnot, reagor2016quantum, wang2016schrodinger, gao2019entanglement,kim2025ultracoherent,milul2023superconducting}. 
These considerations motivate an architectural principle that preserves the advantages of 3D hardware while avoiding uncontrolled growth in local complexity.

A natural approach is modularization, e.g., a large processor is assembled from repeatable unit cells connected through a structured inter-module communication network~\cite{eriksson2024universal, vermersch2017quantum, kimble2008quantum, nickerson2013topological, brown2016co, cirac1999distributed, breuckmann2016constructions,monroe2014large, jiang2007distributed}. 
Each module contains a bounded set of local quantum resources and control interfaces, while communication between modules is mediated through a limited number of external ports. 
Such an architecture distributes fabrication, assembly, calibration, and communication requirements across repeatable hardware units and naturally supports hierarchical control and parallel operation. 
These advantages are not automatic: they depend critically on the interconnect topology and on whether the communication network introduces excessive routing distance, resource concentration, or scheduling overhead.

In this work, we develop an architecture-level framework for modular 3D bosonic quantum processors based on coupled-cavity modules~\cite{kim2025ultracoherent,milul2023superconducting,gao2019entanglement,reagor2016quantum,wang2016schrodinger}.
 The framework tracks where the costs of scale emerge and how those costs are redistributed, when fixed-size bosonic modules are connected into a larger processor.
 Using graph-theoretic proxies~\cite{grinberg2023introduction,bondymurty1982}, we place interconnect count, module connectivity, communication distance, routing concentration, scheduling complexity, and transfer exposure on a common scaling basis and compare their behavior across nearest-neighbor, hub-based, and hybrid architectures.
This comparison shows that the different scaling costs are not independent; reducing one burden can shift or concentrate the cost elsewhere in the architecture. 
The central outcome is therefore not a universally optimal topology, but an architecture-level map of scaling tradeoffs and cost redistribution that identifies where distinct architectural bottlenecks emerge as the processor grows. 
Scalable modular architectures instead require architecture--hardware--workload co-design, with the preferable interconnect strategy depending on physical link quality, bandwidth and multiplexing capacity, and workload locality.

Throughout this comparison, we use uniform all-to-all communication as a topology-neutral reference workload rather than as a universal model of algorithmic traffic; realistic workloads can exhibit strongly nonuniform and locality-dependent intermodule communication~\cite{rached2025characterizing}. 
We likewise treat the interconnect graph as the installed physical connectivity. 
Tunable couplers may activate different interactions within this hardware graph~\cite{kounalakis2018tuneable}, while more general microwave routers or switching fabrics can realize time-dependent effective connectivity and parallel communication paths~\cite{zhou2023realizing,wu2024modular}. 
Their benefit nevertheless remains constrained by the available ports, switching latency, and concurrent transfer capacity. 
The present framework therefore provides a static hardware baseline that can be extended to workload-aware and dynamically reconfigurable architectures.

This article is organized as follows. 
Section~\ref{sec:modular_architecture} introduces an ultra-coherent superconducting-cavity multi-qudit platform as the basic computational module for architectural scaling. 
Section~\ref{sec:architecture_metrics} gives the physical interpretation of graph proxies for the three representative architectures. 
Section~\ref{sec:scaling_laws} develops the corresponding scaling laws and compares design regimes asymptotically.
Section~\ref{sec:case_study} examines a \(3\times3\) processor as a finite-size example and studies a larger hierarchical construction in which fixed-size module groups are assembled into a structured processor. 
Together, these results provide a compact framework for evaluating modular 3D bosonic architectures and identifying the device-level parameters needed for quantitative design, as discussed in Section~\ref{sec:practical_constraints}.

\begin{figure*}
    \centering
    \includegraphics[width=1.0\linewidth, trim=0cm 0.4cm 0cm 0.5cm, clip]{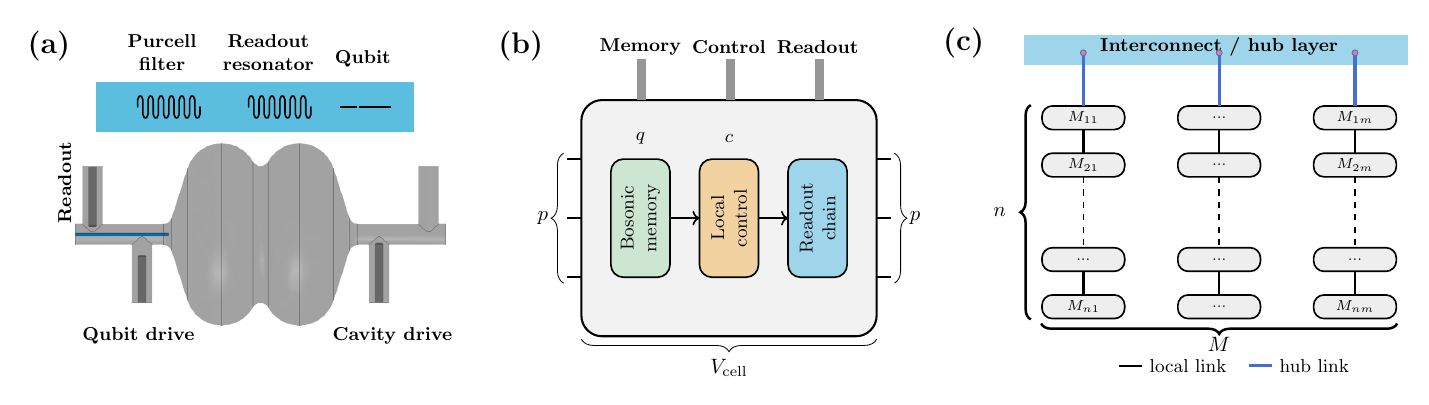}
    \caption{
    Physical-to-architectural mapping for a modular bosonic QPU built from TBMs. 
    (a) Physical schematic of a TBM comprising an two-cell elliptical SRF cavity, ancillary transmon-based control and readout components, and external drive and readout interfaces. 
    (b) Graph-level abstraction of one TBM, with local memory, control, and readout resources grouped into a module vertex. 
    (c) Example of a hybrid communication graph containing \(M\) linear clusters of \(n\) modules.}
    \label{fig:physical_to_architecture}
\end{figure*}

\section{Modular Architecture}\label{sec:modular_architecture}
In this section, we define the elementary hardware unit of the proposed modular processor and state the assumptions used to analyze interconnect scaling. 
The architecture is not modeled as a uniform network of abstract qubits~\cite{rafiee2012stationary,daiss2021quantum}. 
Instead, it is assembled from repeatable, high-performance bosonic modules that provide local memory, control, and readout. 
The central question is how to connect these physically constrained modules while preserving their local functionality and keeping communication and control overhead manageable as the processor scales.

\subsection{two-mode bosonic module}
As illustrated in Fig.~\ref{fig:physical_to_architecture}(a), we adopt a two-mode bosonic module (TBM) as the basic unit~\cite{kim2025ultracoherent}. 
Each TBM consists of a coupled two-cell elliptical SRF cavity structure whose hybridized bosonic modes provide the local degrees of freedom for quantum-information storage and processing, together with a high-coherence ancillary nonlinear element~\cite{bal2024systematic} that enables control, state preparation, and readout.
Relevant capabilities include long-lived multimode storage, mode-selective control, and operations involving accessible bosonic modes. 
We model each TBM as a bosonic computational unit that integrates local quantum-information storage, processing, and ancilla-mediated control.
Quantum information can be stored and processed locally, while selected states or operations are routed through an inter-module communication network when nonlocal interactions are required. 

Fig.~\ref{fig:physical_to_architecture}(b) represents each TBM by a bounded set of local resources and external interfaces at the architectural level. 
We denote by \(q\) the effective local quantum-information capacity under a specified encoding, by \(c\) the number of local control or coupling primitives available for intra-module operations, and by \(p\) the maximum number of external communication interfaces through which the module can connect to the inter-module network. 
Each module also occupies a finite footprint or volume \(V_{\mathrm{cell}}\), and requires \(l_{\mathrm{ctrl}}\) local control lines and \(l_{\mathrm{ro}}\) local readout lines.

These quantities define the fixed local hardware cost of one module. 
In this work, we assume that \(q\), \(c\), \(p\), \(V_{\mathrm{cell}}\), \(l_{\mathrm{ctrl}}\), and \(l_{\mathrm{ro}}\) do not scale with the total number of modules \(N\). 
The processor scaling is governed by the number of repeated modules and by the topology of the inter-module communication network. 
This assumption isolates interconnect-level scaling; it does not describe possible redesigns in which the module capacity, port count, or local control resources change with processor size.

\subsection{Inter-module communication layer}
A processor built from many TBMs requires an interconnect layer that supports operations beyond a single local node.
Fig.~\ref{fig:physical_to_architecture}(c) illustrates such an interconnection, in which \(M\) linear clusters, each containing \(n\) TBMs, are connected through a shared hub layer.
Depending on the hardware realization, this layer may be implemented using direct couplers, shared hub modes, interposers, vertical communication channels, or sparse local routing networks~\cite{gold2021entanglement, niskanen2007quantum, renger2026superconducting, kosen2022building, liao2026breaking, kivlichan2018quantum, arute2019quantum}. 
In this work we do not prescribe a specific coupling mechanism. 
Instead, we represent the interconnect by an abstract communication graph that captures which modules can communicate directly or through shared routing resources.

The interconnect is treated as a constrained architectural resource, not as an ideal communication channel. 
Adding ports, links, or shared modes can improve reachability, but it can also increase packaging complexity, insertion loss, parasitic mode participation, spectral crowding, calibration burden, and control overhead. 
Shared pathways can further limit parallel operation when multiple transfers or mediated interactions compete for the same resource.
The design objective is not simply to maximize connectivity, but to supply the nonlocal communication required by the intended workload while limiting the associated physical and operational cost.
Since each TBM already provides long-lived storage and local control, the processor-level performance may be limited by the cost of moving quantum information between modules rather than by the local memory itself. 
The architecture should therefore be evaluated by how the interconnect distributes communication distance, shared-resource exposure, routing concentration, and control overhead across the processor.

\subsection{Architectural abstraction}
The architectural model separates the fixed resources contained within each TBM from the communication resources used to connect different modules.
Local memory capacity, control primitives, readout capability, physical footprint, and external port limits are assigned to the modules themselves.
By contrast, communication distance, routing burden, shared-resource exposure, and potential contention are determined by the inter-module network.

This separation allows processors with the same local modules but different interconnects to be compared on a common basis.
At the same time, the graph representation is intentionally limited: it records the availability and organization of communication pathways, but does not by itself specify coupling strengths, transfer fidelities, coherence losses, gate times, or frequency-allocation constraints.
Such device-dependent quantities must be supplied separately when a particular physical implementation is considered.

\section{Architectural metrics}\label{sec:architecture_metrics}
Motivated by prior studies of multimode control, communication topology, and routing in modular quantum systems~\cite{ma2021quantum,naik2017random,linke2017experimental, beals2013efficient,chou2018deterministic}, we organize the comparison around four architecture-level quantities: structural connectivity, communication distance, routing concentration, and control complexity. 
These quantities are topological proxies for comparing interconnect strategies; each proxy captures a different way in which interconnect topology can impose physical or operational costs.

\subsection{Physical interpretation of the graph proxies}
Each topological proxy is associated with a concrete physical cost, allowing hardware-level constraints to be propagated into architecture-level scaling behavior.

\emph{Structural connectivity} specifies which module-module or module-resource pairs are directly linked. 
It is constrained by the module port budget and is quantified by edge count, module degree, and the degree of explicitly modeled interconnect resources. 
Connectivity indicates the scale of the required communication interface, but not its full wiring or packaging cost~\cite{gold2021entanglement, wu2024modular, niu2023low}.

\emph{Communication distance} specifies how many graph steps separate two computational modules. 
We use shortest-path distance as a proxy for the number of sequential transfers or mediated interactions required for nonlocal operations. 
Latency and fidelity depend on the physical rates and errors of the links along those paths~\cite{niu2023low, cowtan2019qubit}.

\emph{Routing concentration} specifies whether communication paths are distributed across the processor or concentrated on a small number of modules, gateways, or shared resources. 
Concentration is associated with congestion, bandwidth demand, shared loss exposure, crosstalk, calibration sensitivity, and single-resource failure risk~\cite{newman2005measure,renger2026superconducting}.

\emph{Control complexity} specifies the difficulty in scheduling and coordinating the use of the interconnect. 
We use edge coloring as a restricted proxy for link-conflict scheduling, together with routing concentration as an indicator of shared-resource demand. 
These proxies do not describe the full pulse-control or calibration problem, but they show when graph structure forces serialization or centralized coordination~\cite{guerreschi2018two}.

Wiring and packaging depend on ports, couplers, feedthroughs, routing area, multiplexing, and control or readout access. 
Loss and crosstalk depend on insertion loss, parasitic participation, unwanted hybridization, frequency crowding, and interference introduced by links, gateways, and shared routing elements.
Therefore, these parameters are treated as physical consequences of these graph proxies rather than as independent topological metrics.

\subsection{Graph metrics for interconnect scaling}
We represent a modular processor and its explicitly modeled interconnect resources by a finite, connected, undirected, simple augmented graph~\cite{caleffi2024distributed}
\begin{equation}
    G=(V_{\mathrm M}\cup V_{\mathrm I},E).
\end{equation}
The set \(V_{\mathrm M}\) contains computational module vertices, while \(V_{\mathrm I}\) contains auxiliary interconnect vertices such as hubs, routers, or mediated communication modes. 
An edge \(e\in E\) represents an available communication link, either directly between modules or between a module and an interconnect resource. 
For architectures described entirely by direct module-to-module links, \(V_{\mathrm I}=\varnothing\).

The number of computational modules is \(N=|V_{\mathrm M}|.\)
Under the fixed-module assumption, the nominal aggregate local quantum-information capacity is \(Q(N)=Nq,\) before encoding, communication, or error-correction overhead. 
The metrics below are primary graph quantities, providing quantitative descriptions of the four proxies. 

\emph{Interconnect count:} We define the total number of communication links as
\begin{equation}
    W=|E|=\frac{1}{2}\sum_{v\in V_{\mathrm M}\cup V_{\mathrm I}}\mathrm{deg}(v),
\end{equation}
where \(\mathrm{deg}(v)\)  denotes the number of communication links incident on a vertex \(v\).
The edge count \(W\) is a coarse proxy for interconnect size, but not a complete wiring or packaging cost.

The average number of external communication interfaces used by each module is
\begin{equation}
    \overline{K}_{\mathrm M}=\frac{1}{N}\sum_{v\in V_{\mathrm M}}\mathrm{deg}(v),
\end{equation}
and the maximum module degree is
\begin{equation}
    \Delta_{\mathrm M}=\max_{v\in V_{\mathrm M}}\mathrm{deg}(v).
\end{equation}
Since each TBM has at most \(p\) external communication ports, \(\overline{K}_{\mathrm M}\leq \Delta_{\mathrm M}\leq p.\)
For architectures with explicit interconnect vertices, we also define
\begin{equation}
    \Delta_{\mathrm I} =
    \begin{cases}
        \displaystyle\max_{v\in V_{\mathrm I}}\mathrm{deg}(v), & V_{\mathrm I}\neq\varnothing, \\
        0, & V_{\mathrm I}=\varnothing .
    \end{cases}
\end{equation}
\(\overline{K}_{\mathrm M}\) and \(\Delta_{\mathrm M}\) quantify module-level connectivity burden, while \(\Delta_{\mathrm I}\) quantifies degree concentration at shared interconnect resources. 

\emph{Communication distance:} We characterize nonlocal communication by the average shortest-path distance between computational modules,
\begin{equation}
    \bar{\ell} = \frac{1}{N(N-1)} \sum_{\substack{u,v\in V_{\mathrm M},\\u\neq v}} d_G(u,v),
\end{equation}
where \(d_G(u,v)\) is measured in the full graph \(G\). 
Paths may therefore pass through auxiliary interconnect vertices. 
The quantity \(\bar{\ell}\) is a proxy for the average number of sequential communication steps required for module-to-module operations.

\emph{Routing concentration:} We quantify routing concentration using the normalized maximum shortest-path betweenness over module-to-module traffic
\begin{equation}
\widetilde{B}=\frac{1}{N(N-1)}\max_{v\in V_M\cup V_I}\sum_{\substack{s,t\in V_M\\ s\neq t,\;s,t\neq v}}\frac{\sigma_{st}(v)}{\sigma_{st}} .
\end{equation}
Here, \(\sigma_{st}\) is the number of shortest paths between modules \(s\) and \(t\), and \(\sigma_{st}(v)\) is the number of those paths that pass through \(v\). 
The normalization is taken over the \(N(N-1)\) ordered pairs of computational modules.

\emph{Scheduling complexity:} We use the edge chromatic number as a proxy for link-conflict scheduling,
\begin{equation}
    C=\chi'(G).
\end{equation}
This corresponds to the minimum number of independent link classes required to form a proper edge coloring, ensuring that no two links incident on the same vertex share the same class.
In an architecture where a single module port can only participate in one operation at a time, two links sharing a vertex cannot be active simultaneously. 
A proper edge coloring is necessary to partition the edge set $E$ into $C$ distinct, conflict-free link layers. 
Physical implementation details, such as frequency multiplexing, pulse constraints, or non-local crosstalk, can further modulate the actual schedule depth, but $C$ establishes the baseline routing complexity dictated solely by the hardware topology.

Together, \(W\), \(\overline{K}_{\mathrm M}\), \(\Delta_{\mathrm M}\), \(\Delta_{\mathrm I}\), \(\bar{\ell}\), \(\widetilde{B}\), and \(C\) form the graph-metric set used below. 
They provide an architecture-level description of link count, module port usage, shared-resource degree, communication distance, routing concentration, and link-conflict scheduling. 

\begin{figure*}
    \centering
    \includegraphics[width=1.0\linewidth, trim=0.0cm 0.4cm 0.0cm 0.3cm, clip]{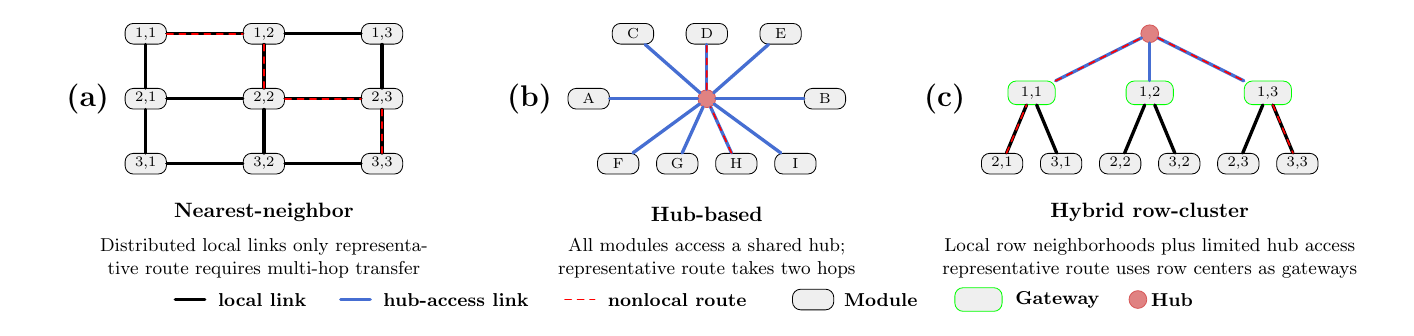}
    \caption{Comparison of three interconnect topologies for the same bosonic processor with 9 computational modules. 
    (a) Nearest-neighbor grid with only local links; a representative nonlocal route requires multiple hops. 
    (b) Hub-based architecture with a central hub connecting to all modules; the representative route uses two module-hub links. 
    (c) hybrid row-cluster architecture that retains local row-chain connectivity while designating each row-center module as a gateway to the central hub.}
    \label{fig:topology_comparison}
\end{figure*}

\section{Scaling laws}\label{sec:scaling_laws}
Using the architectural metrics defined above, we compare three representative interconnect strategies for assembling many TBMs: nearest-neighbor, hub-based, and hybrid architectures. 
This comparison identifies how different graph structures redistribute the costs of scale, including communication distance, link count, shared-resource exposure, routing concentration, and scheduling overhead~\cite{zhou2023realizing, pietikainen2024strategies,monroe2014large,ang2024arquin, mollenhauer2025high,singh2025modular}.
Although many other communication graphs are possible, these three architectures capture the central tradeoff between distributed local connectivity and centralized routing resources. 
Figure~\ref{fig:topology_comparison} illustrates the three realizations, each containing the same number of TBMs (\(N=9\)), and thereby provides a minimal setting for nontrivial comparison of their graph metrics. 
Detailed derivations of these metrics are given in Appendices~\ref{app:nn_metrics}--\ref{app:hybrid_metrics}.

\subsection{Nearest-neighbor lattice}
A nearest-neighbor lattice represents the most local interconnect architecture considered here. 
In an open \(n\times n\) square array, each module communicates with its horizontal and vertical neighbors.  

For \(N=n^2\) modules, the number of physical links is 
\begin{equation}
    W_{\mathrm{NN}}=2N-2\sqrt{N},
\end{equation}
and the maximum module degree is 
\begin{equation}
    \Delta_{M,\mathrm{NN}}=4.
\end{equation} 
The architecture has bounded local connectivity and does not require any dedicated shared routing resources. 
It is therefore attractive from a hardware-integration perspective: each module only needs a small, fixed number of local interfaces, independent of system size.

This locality, however, does not remove the cost of nonlocal communication. 
It relocates that cost into repeated transport through the computational array itself. 
A message or quantum state routed between two distant modules must traverse a sequence of nearest-neighbor hops. 
For uniformly chosen source--destination pairs, the mean shortest-path distance on the square lattice is
\begin{equation}
    \bar{\ell}_{\mathrm{NN}}=\frac{2}{3}\sqrt{N},    
\end{equation}
while the worst-case distance between opposite corners is \(\ell_{\rm NN}^{\rm max}=2(\sqrt{N}-1)\). 
Both typical and worst-case communication distances grow as \(\sqrt{N}\), even though the local degree remains constant.

This scaling has several engineering consequences. 
First, any loss, infidelity, or conversion penalty incurred per hop accumulates over the full route. 
If each hop has effective transmission \(\eta_0<1\), a path of length \(d\) has transmission \(\eta_0^d\). 
Assuming \(1-\eta_0\ll1\), the mean path-loss exposure scales approximately as
\begin{equation}
\overline{\mathcal{L}}_{\mathrm{NN}}\simeq\frac{2}{3}(1-\eta_0)\sqrt{N}.
\end{equation}
It is therefore required that the effective per-hop loss improves roughly as \(N^{-1/2}\) to maintain a fixed typical end-to-end transmission as the lattice grows, unless additional architectural mechanisms reduce the routing distance.

Second, communication traffic is carried by the computational modules themselves.
Although there is no centralized router and no auxiliary shared-hardware bottleneck, intermediate modules must participate in transporting information between distant pairs.
The routing burden is therefore distributed across the array, with modules near the center generally lying on more shortest paths than boundary modules. 
Asymptotically, the normalized traffic concentration on the busiest module decreases with the system size~\cite{lampo2021multiple,kumar2019betweenness}
\begin{equation}
    \widetilde{B}_{\mathrm{NN}}\simeq\frac{9}{4}\frac{\overline{B}_{\rm NN}}{N(N-1)}\sim \frac{3}{2\sqrt{N}}.
\end{equation}
Here, \(\overline{B}_{\rm NN}=(N-1)(\bar \ell_{\rm NN}-1)\) is the mean shortest-path betweenness.
Although nonlocal communication requires increasingly long multihop routes, the associated transit burden remains distributed across the computational lattice rather than becoming concentrated at a single shared routing resource.

Finally, the physical links of the lattice are locally easy to schedule. 
Because the square lattice is bipartite and has maximum degree four, meaning that each module is connected to at most four neighboring modules, its edge chromatic number is
\begin{equation}
    \chi'(G_{\mathrm{NN}})=4.
\end{equation}
Equivalently, all nearest-neighbor links can be partitioned into four conflict-free activation layers under a one-link-per-module constraint. 
This constant-depth link-coloring property describes only one complete activation of the physical links. 
Nonlocal operations still require repeated activations over many hops, so their latency and accumulated exposure scale with the routing distance.

Overall, the nearest-neighbor lattice makes a clear architectural tradeoff. 
It minimizes local connectivity and avoids dedicated shared routing hardware, but it transfers the scaling burden to multi-hop transport: longer paths, accumulated loss, increased latency, and routing traffic through computational modules. 

\begin{table*}[t]
\footnotesize
\centering
\caption{Leading-order asymptotic topological proxies for nearest-neighbor (NN), hub-based, and balanced gateway--hub hybrid modular interconnects. Module-level degree metrics are separated from auxiliary shared-resource degree. Here \(\bar{b}=\Delta_{\mathrm I}/N\) is the normalized shared-resource burden.}\label{tab:architecture_comparison}
\begin{ruledtabular}
\begin{tabular}{l c c c c c c c c l}
&\(W\) &\(\overline{K}_{\mathrm M}\) &\(\Delta_{\mathrm M}\) &\(\Delta_{\mathrm I}\) &\(\bar b\)&\(\chi'(G)\) &\(\bar{\ell}\)
&\(\widetilde B\)
&\parbox[t]{8.0cm}{\textbf{Design interpretation}} \\[0.5ex]
\hline
\textbf{NN}\footnote{NN values assume an open square grid with \(N=n^2\) modules.}
&\(\mathcal O(N)\) &\(4\) &\(4\) &\(0\) &\(0\) &\(4\) &\(\frac{2}{3}\sqrt N\)
&\(N^{-1/2}\)
&\parbox[t]{8.0cm}{Bounded local connectivity and no shared router; nonlocal communication accumulates distance, latency, and loss.}\\
\hline
\textbf{Hub}
&\(N\) &\(1\) &\(1\) &\(N\) &\(1\) &\(N\) &\(2\)
&\(1\)
&\parbox[t]{8.0cm}{Constant graph distance; fan-out, scheduling, bandwidth, and calibration burden concentrate at the shared resource.}\\[3mm]
\hline
\textbf{Hybrid}\footnote{Hybrid values in this row assume \(M=N/n\) linear clusters, with the balanced choice \(n=M\) and centrally placed gateways.}
&\(\mathcal O(N)\) &\(\mathcal O(1)\) &\(\mathcal O(1)\) &\(\sqrt N\) &\(N^{-1/2}\) &\(\sqrt N\) &\(\sim \frac{1}{2}\sqrt N\)
&\(\tfrac{N-n}{N-1}\)
&\parbox[t]{8.0cm}{Reduces hub fan-out relative to a direct hub; retains local multihop access and depends on workload locality.}
\end{tabular}
\end{ruledtabular}
\end{table*}

\subsection{Hub-based interconnect}
A hub-based architecture takes the opposite approach from the nearest-neighbor lattice. 
Each of the \(N\) computational modules connects to a single shared routing resource, so the routing graph is a star. 
This gives the smallest possible module-level connectivity: every computational module has degree one, and the total number of module--hub links is
\begin{equation}
    W_{\mathrm{hub}}=N.
\end{equation}
The cost is transferred to the auxiliary routing resource with maximum hub degree
\begin{equation}
    \Delta_{I,\mathrm{hub}}=N.
\end{equation}
The hub architecture is therefore locally economical at the module level, but maximally centralized at the shared interconnect. 

The main advantage of this centralization is constant graph distance. 
Since any two distinct modules communicate through the route \(u\rightarrow h\rightarrow v\), the mean and worst-case module-to-module distances are both
\begin{equation}
    \bar{\ell}_{\mathrm{hub}}=\ell_{\rm hub}^{\rm max}=2.
\end{equation}
If each module--hub link has effective transmission \(\eta_0\) and the internal hub operation has efficiency \(\eta_h\), every module-to-module route has transmission \(\eta_0^2\eta_h\). The corresponding path-loss probability is
\begin{equation}
    \overline{\mathcal{L}}_{\mathrm{hub}}=1-\eta_0^2\eta_h.
\end{equation}
Unlike the nearest-neighbor lattice, the hub avoids the accumulated multi-hop loss at the graph level.
This conclusion assumes, however, that \(\eta_0\) and \(\eta_h\) remain independent of hub fan-out, physical link length, and traffic load.

The routing burden is also completely concentrated. 
For ordered module pairs, every shortest path passes through the hub, so the hub betweenness is \(B_{\mathrm{hub}}(h)=N(N-1)\sim N^2\), whereas every computational module has zero shortest-path betweenness.
The hub therefore carries the full module-to-module transit load.
Normalizing by the number of ordered computational-module pairs gives
\begin{equation}
    \widetilde{B}_{\rm hub}=\frac{B_{\rm hub}(h)}{N(N-1)}=1,
\end{equation}
showing that every module-to-module shortest path traverses the same intermediate resource.

This centralization also appears in link scheduling. 
No two links can be activated simultaneously under a one-operation-per-resource constraint because every module--hub edge is incident on the same hub vertex. 
The edge chromatic number is therefore
\begin{equation}
    \chi'(G_{\mathrm{hub}})=N.
\end{equation}
Activating every module--hub link once requires \(N\) conflict-free layers in this topology-only model. 
Constant-depth activation becomes possible only if the hub supports a multiplexing capacity that scales with system size, for example through parallel modes, frequency channels, switching elements, or replicated subresources~\cite{zhou2023realizing,roslund2014wavelength,ruskuc2025multiplexed}.

Overall, the hub architecture trades multi-hop transport for centralized shared hardware. 
It gives constant graph distance, simple module interfaces, and size-independent per-transfer loss in an idealized fixed-quality model. 
Its scaling burden is instead concentrated in the hub: fan-out, bandwidth, multiplexing capacity, isolation, calibration, and aggregate traffic must all scale with the number of connected modules.

\subsection{Hybrid clustered interconnect}
A hybrid architecture interpolates between a fully local lattice and a fully centralized hub. 
We consider \(M=N/n\) local clusters, each containing \(n\) computational modules. 
Each cluster has a designated gateway module \(g_i\), and the gateways are connected by a higher-level interconnect \(H\). 
The gateway remains a computational module, but also serves as the communication node between its local cluster and the higher-level routing network.

This construction separates local module connectivity from shared-resource connectivity. 
The total number of links is
\begin{equation}
    W_{\mathrm{hyb}}=M(n-1)+|E_H|,
\end{equation}
where \(M(n-1)\) counts the links inside the linear clusters and \(|E_H|\) counts the higher-level interconnect links. 
The mean computational-module degree is
\begin{equation}
    \overline{K}_{M,\mathrm{hyb}}=2-\frac{2}{n}+\frac{\bar d_{H,g}}{n},
\end{equation}
where \(\bar d_{H,g}\) is the mean number of higher-level links incident on a gateway. 
For bounded gateway degree, the module-level connectivity remains local and size independent.

For the gateway--hub realization, all \(M\) gateways connect to one shared hub. In this case,
\begin{equation}
    |E_H|=M,\quad \Delta_{I,\mathrm{hyb}}=M=\frac{N}{n}.
\end{equation}
Compared with a direct module--hub star, where the hub degree is \(N\), clustering reduces the shared-resource fan-out by a factor of \(n\). 
In the balanced case \(n=M=\sqrt{N}\), the hub degree scales as \(\sqrt{N}\) rather than \(N\). 
The price is that modules must first reach their local gateway through a multi-hop local path.

The mean communication distance reflects this two-level structure. 
If denoting \(p_{\mathrm{s}}=(n-1)/(N-1)\) as the probability that a destination lies in the same cluster and \(p_{\mathrm{d}}=(N-n)/(N-1)\) as the probability that it lies in a different cluster, we obtain
\begin{equation}
    \bar{\ell}_{\mathrm{hyb}}=p_{\mathrm{s}}\bar{\ell}_{\mathrm{loc}}(n)+p_{\mathrm{d}}\left[2\bar{\ell}_{g}(n)+\bar{\ell}_{H}(M)\right].
\end{equation}
Here, \(\bar{\ell}_{\mathrm{loc}}(n)=(n+1)/3\) is the mean distance between two modules in the same linear cluster, \(\bar{\ell}_{g}(n)\) is the mean distance from a module to its gateway, and \(\bar{\ell}_{H}(M)\) is the mean distance between gateway terminals in the higher-level network. 
In the balanced case \(n=M=\sqrt{N}\), the dominant contribution comes from local access to the gateway: the mean distance scales approximately as \(\frac{1}{2}\sqrt{N}\) for a centrally placed gateway, and as \(\sqrt{N}\) for an endpoint gateway. 
The hybrid reduces the hub fan-out but does not eliminate multi-hop local transport.

\begin{figure*}
    \centering
    \includegraphics[width=1.0\linewidth, trim=0.0cm 0.4cm 0.0cm 0.0cm, clip]{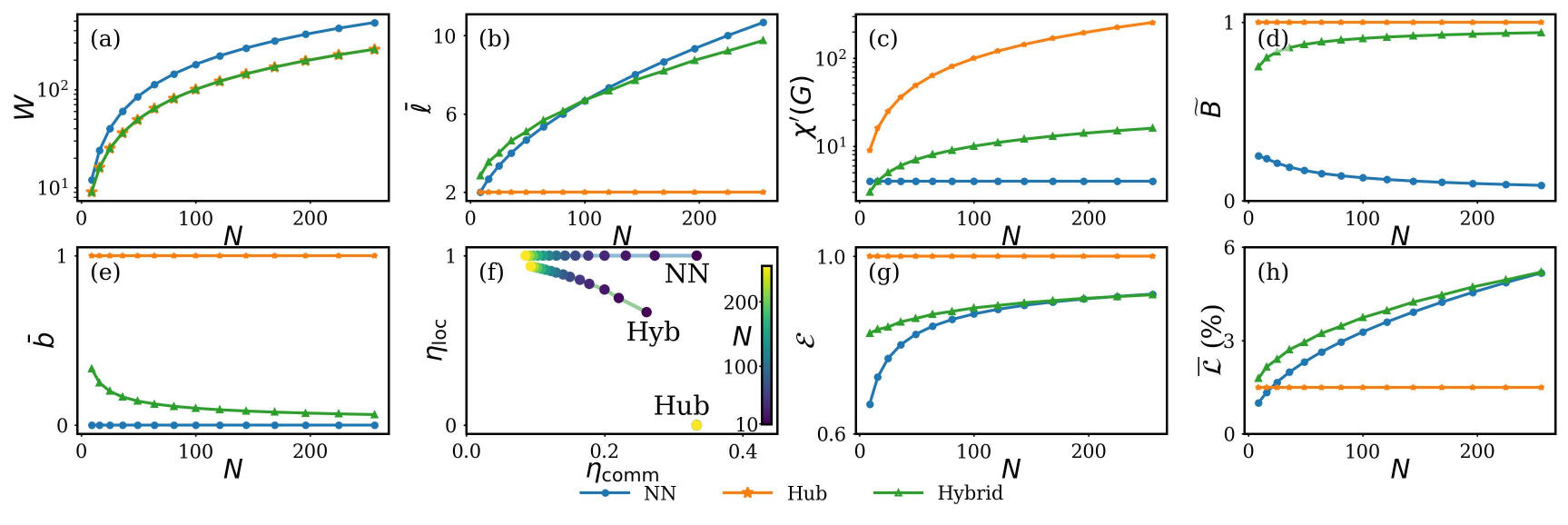}
    \caption{Architecture-level scaling comparison for the nearest-neighbor (NN), hub, and balanced hybrid interconnects.
    (a) Link count \(W\), characterizing the total hardware-connectivity requirement.
    (b) Mean shortest-path distance \(\bar{\ell}\) between computational modules, setting the baseline number of transfer steps required for communication.
    (c) Link-conflict scheduling depth \(C=\chi'(G)\) under exclusive-endpoint activation, for which links sharing a module or routing resource cannot operate simultaneously.
    (d) Normalized maximum shortest-path betweenness \(\widetilde{B}\), characterizing the concentration of shortest-path traffic through intermediate vertices.
    (e) Normalized shared-resource burden \(\bar{b}=\Delta_r/N\), given by the maximum auxiliary-resource degree relative to system size.
    (f) Communication--decentralization plane defined by the distance-based score \(\eta_{\mathrm{comm}}=1/(1+\bar{\ell})\) and decentralization score \(\eta_{\mathrm{loc}}=1-\bar{b}\); marker color and size indicate increasing \(N\).
    The hub points overlap because both coordinates are independent of \(N\).
    (g) Composite architectural exposure \(\mathcal{E}=1-\eta_{\mathrm{comm}}\eta_{\mathrm{loc}}\), which increases with communication distance, shared-resource centralization, or both.
    (h) Illustrative transfer loss \(\overline{\mathcal{L}}\) associated with the mean path length for per-link efficiency \(\eta_0=0.995\), showing the accumulation of small per-link losses over multihop communication.
    The quantities \(\eta_{\mathrm{comm}}\), \(\eta_{\mathrm{loc}}\), and \(\mathcal{E}\) in panels (f)--(g) are visualization indicators rather than physical efficiencies or fidelities and should not be interpreted as direct operational-performance measures.}
    \label{fig:scaling_comparison}
\end{figure*}

The routing concentration shows the same tradeoff. 
In the gateway--hub hybrid, intracluster traffic stays inside the local cluster, but every intercluster route passes through the shared hub. 
Under uniform all-to-all ordered traffic, the normalized concentration is
\begin{equation}
    \widetilde{B}_{\rm hyb}=\frac{N-n}{N-1}.
\end{equation}
Increasing the cluster size reduces the hub degree from \(N\) to \(N/n\), but it does not necessarily remove traffic concentration. 
When \(n\ll N\), most ordered module pairs lie in different clusters, so most all-to-all traffic still traverses the higher-level hub.

The link-scheduling cost is likewise set by the most concentrated part of the routing graph. 
For the gateway--hub star, the full graph is bipartite and the hub has degree \(M\), giving the exact scheduling depth
\begin{equation}
    \chi'(G_{\mathrm{hyb}})=M=\frac{N}{n}.
\end{equation}
In the balanced case, this becomes \(\chi'(G_{\mathrm{hyb}})=\sqrt{N}\). 
Therefore, clustering reduces the \(N\)-layer scheduling burden of a direct hub, but constant-depth activation still requires hub multiplexing capacity that grows with the number of clusters.

We separate path loss into local and higher-level contributions. 
For the balanced gateway--hub hybrid with a centrally placed gateway, the leading behavior of the mean loss in the weak regime is
\begin{equation}
    \overline{\mathcal{L}}_{\mathrm{hyb}}\simeq\frac{\sqrt{N}}{2}(1-\eta_0)+2(1-\eta_H)+(1-\eta_h),
\end{equation}
in which we assume local cluster links have efficiency \(\eta_0\), gateway--hub links have efficiency \(\eta_H\), and the hub has internal efficiency \(\eta_h\).
The hybrid architecture reduces the normalized hub burden, but the local multi-hop exposure still grows with the maximum shortest-path distance between any two modules within the same cluster.
Maintaining bounded local loss as \(N\) grows requires the local per-hop loss to improve roughly as \(N^{-1/2}\), unless the cluster size, gateway placement, or higher-level topology is changed.

Overall, the hybrid architecture is best viewed as a tunable co-design space rather than a universal scaling solution. 
Larger clusters reduce shared-resource fan-out and scheduling pressure at the higher level, but increase local access distance and accumulated local loss. 
Smaller clusters reduce local transport distance, but increase the number of gateways and the burden placed on the higher-level interconnect. 
The architecture therefore trades local-link quality, gateway placement, cluster size, hub capacity, and workload locality against one another explicitly.

\subsection{Asymptotic comparison and design regimes}
Table~\ref{tab:architecture_comparison} and Figure~\ref{fig:scaling_comparison} summarize the leading topology-level tradeoffs among the three architectures; detailed derivations are provided in Appendices~\ref{app:nn_metrics}--\ref{app:hybrid_metrics}. 
The uniform all-to-all traffic model used here should be viewed as a topology-independent baseline rather than a representative workload. 
It removes communication locality and therefore understates the main advantage of hybrid architectures, while naturally highlighting the short-path benefit of a hub. 
At the same time, it exposes the hub's concentrated routing and scheduling burden. 
The asymptotic comparison should therefore be interpreted as baseline scaling comparisons, not as a workload-independent ranking of the three architectures.

The primary asymptotic metrics in Figs.~\ref{fig:scaling_comparison}(a)--~\ref{fig:scaling_comparison}(d) quantify link count \(W\), mean communication distance \(\bar \ell\), link-conflict scheduling depth \(C\), and routing concentration \(\overline{B}\). 
To make the distance--centralization tradeoff more transparent, Figs.~\ref{fig:scaling_comparison}(e)--~\ref{fig:scaling_comparison}(g) introduce the normalized shared-resource burden \(\bar b=\Delta_{\rm I}/N\), the communication and locality coordinates \(\eta_{\rm comm}=1/(1+\bar{\ell})\) and \(\eta_{\rm loc}=1-\bar b\), and the illustrative exposure \(\mathcal{E}=1-\eta_{\rm comm}\eta_{\rm loc}\).
Here, \(b\) measures reliance on a high-degree shared routing resource, while the communication--locality plane displays the balance between short paths and distributed connectivity. 
The quantity \(\mathcal{E}\) compactly highlights whether the dominant exposure arises from multihop routing, shared-resource centralization, or both; it is intended as a visualization aid rather than a physical error probability.
Figure~\ref{fig:scaling_comparison}(h) separately translates \(\bar \ell\) into an illustrative accumulated path loss \(\overline{\mathcal{L}}\) for a fixed per-link efficiency \(\eta_0\). 

\begin{figure*}[t!]
    \centering
    \includegraphics[width=1.0\linewidth, trim={0cm 0.4cm 0cm 0.4cm}, clip]{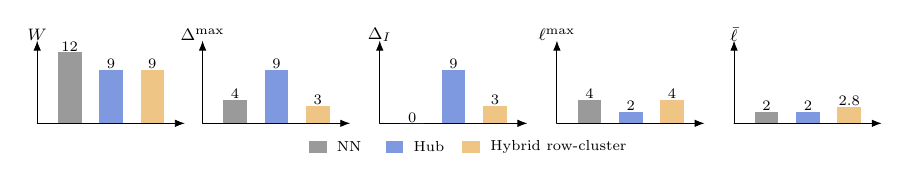}
    \caption{Topological proxy comparison for three \(3\times3\) modular interconnects: nearest-neighbor (NN), global hub, and hybrid row-cluster. 
    Panels show total number of links \(W\), maximum graph degree \(\Delta^{\rm max}\), shared-hub burden \(\Delta_{\rm I}\), worst-case module-to-module path length \(\ell^{\rm max}\), and mean path length \(\bar\ell\).}
    \label{fig:case_study} 
\end{figure*}

The nearest-neighbor lattice realizes the most distributed regime. 
It relies only on local connectivity, maintains bounded module degree and constant scheduling depth, and eliminates auxiliary shared-resource burden. 
Its principal engineering advantages are therefore hardware regularity, bounded port count, and distributed routing load. 
This regime is particularly well suited to workloads with strong communication locality or sufficiently high-fidelity local transfer.

The hub architecture realizes the most centralized regime, but provides the strongest global-connectivity advantage. 
Every module-to-module route has distance two, while each module requires only a single communication interface. 
A hub is therefore attractive when rapid global access is essential and a high-fan-out, high-bandwidth shared routing resource can be engineered.

The hybrid architecture provides a tunable intermediate regime by combining local clusters with a higher-level gateway network. 
Its principal advantage is that it reduces the fan-out and scheduling burden of the shared resource while retaining more efficient nonlocal access than a purely nearest-neighbor architecture. 
This regime is especially attractive when communication exhibits cluster locality and the cluster size and gateway placement can be co-designed with the workload.

Overall, the comparison reveals no topology-only optimum. Nearest-neighbor connectivity prioritizes locality and distributed control, the hub prioritizes minimum communication distance, and the hybrid provides a tunable intermediate regime. The preferred architecture must therefore be selected through co-design of workload locality, cluster size, gateway placement, link quality, routing capacity, scheduling constraints, and shared-resource performance.

\section{Case study: from a \(3\times3\) tile to a hierarchical processor}\label{sec:case_study}
In this section, we examine finite-size examples to show how the same tradeoffs appear in explicit modular layouts. 

\subsection{\(3\times3\) processor}
We first consider the three architectures on a \(N=9\) processor, as illustrated in Figure~\ref{fig:topology_comparison}. 
This is the smallest square processor for which local routing, centralized communication, and gateway-based hierarchy can be compared nontrivially. 
The comparison should be read not as a change in processor layout, but as a relocation of communication resources. 
The corresponding finite-size metrics are summarized in Figure~\ref{fig:case_study}.

The nearest-neighbor layout keeps communication resources fully local. 
This avoids any architectural intervention relative to the baseline, but nonlocal transfers require multi-hop routing across the array.

The hub layout relocates communication to a single shared resource. 
This minimizes routing depth, with \(\ell^{\max}=2\), but with the cost of concentrated access and control at the hub, giving \(\Delta^{\max}=9\).

The hybrid row-cluster layout relocates part of the communication resource to a shared layer. 
By using row-center modules as gateways, it keeps the link count at \(W=9\) and reduces the shared-resource burden to \(\Delta^{\max}=3\), while retaining \(\ell^{\max}=4\).

\begin{table*}[t]
\centering
\caption{Finite-size comparison of a \(3\times3\) tile, a flat \(9\times9\) mesh, and an 81-module hierarchical construction built from nine \(3\times3\) tiles.}
\label{tab:finite_size_comparison}
\begin{ruledtabular}
\begin{tabular}{lccccccc}
Architecture & \(|V|\) & \(|E|\) & \(\overline K\) & \(\Delta^{\max}\) & \(\bar\ell\) & \(\ell^{\max}\) & \(\overline{\mathcal{L}}\)\\
\hline
\(3\times3\) tile 
& 9 & 12 & 2.67 & 4 & 2.0 & 4 & 0.997\%\\
\(9\times9\) flat mesh 
& 81 & 144 & 3.56 & 4 & 6.0 & 16 & 2.952\%\\
\(3\times3\) hierarchical tiles\footnote{The detailed derivations of these metrics are given in Appendix~\ref{app:hierarchical_81}} 
& 81 & 120 & 2.96 & 8 & 4.4 & 8 & 2.179\%
\end{tabular}
\end{ruledtabular}
\end{table*}

\subsection{Extension to a larger hierarchical processor}
The \(3\times3\) tile can also serve as a modular building block for larger processors. To illustrate this finite-size scaling step, Table~\ref{tab:finite_size_comparison} compares an 81-module processor realized either as a flat \(9\times9\) mesh or as a \(3\times3\) array of \(3\times3\) tiles. In the hierarchical construction, each tile preserves its local \(3\times3\) mesh, while the center module of each tile acts as a gateway for inter-tile communication.

The main effect of modularization is to replace a fully extended mesh with a two-level communication structure. Most links remain local within tiles, while only a sparse set of gateway-to-gateway links is used to connect different tiles. As shown in Table~\ref{tab:finite_size_comparison}, this reduces the total number of physical links and lowers the average module degree relative to the flat \(9\times9\) mesh. It also shortens both the mean and worst-case module-to-module distances, because long-range communication can move through the gateway layer rather than diffusing across the full mesh.

These savings are not free. The hierarchy reduces distributed wiring and routing cost by concentrating part of the communication burden onto gateway modules. This is reflected in the increase of \(\Delta^{\max}\) for the hierarchical architecture. Thus, modularization trades many low-level nearest-neighbor connections and long mesh paths for fewer long-range resources that must be engineered with higher connectivity, routing load, and control overhead.

In this sense, the hierarchical processor does not eliminate interconnect cost; it relocates it. The benefit is that most modules retain simple local connectivity, while a small number of designated gateway modules absorb the additional complexity needed for nonlocal communication. This provides a concrete finite-size example of the architectural principle used throughout this work: modular structure can save physical resources and reduce typical communication distance, provided that the cost of the higher-level gateways is explicitly accounted for~\cite{gold2021entanglement, smith2022scaling}.

\section{Design tradeoffs and practical constraints}\label{sec:practical_constraints}
The scaling analysis above treats interconnects as graphs and compares link count, communication distance, routing concentration, and scheduling burden. 
These quantities are useful architectural proxies, but they do not by themselves determine the preferred implementation. 
For a processor built from TBMs, the interconnect must preserve the physical advantages of the module: long coherence, controlled electromagnetic participation, and repeatable local calibration. 
The central design question is therefore not only how many links or hops are required, but how the added communication layer affects congestion, loss, crowding, calibration, and robustness~\cite{pietikainen2024strategies,ang2024arquin,you2024crosstalk,rosenberg20193d,laracuente2025modeling,mollenhauer2025high,singh2025modular}.

\emph{Congestion.}
A short graph path is useful only if the corresponding communication resource is available when the operation is scheduled. 
Nearest-neighbor layouts distribute traffic over many local links, but nonlocal operations require multi-hop routing and may occupy several intermediate resources. 
Hub-based layouts reduce graph distance, but concentrate arbitration and bandwidth demand at a shared element. 
Hybrid layouts provide an intermediate structure: common operations can remain local within a tile or cluster, while selected nonlocal transfers use a higher-level communication layer. 
The relevant figure of merit is therefore not only path length, but the match between interconnect throughput and the local processing rate of the modules.

\emph{Loss.}
Long coherence is a primary motivation for bosonic modules, but each added port, coupler, routing interface, or shared mode can introduce insertion loss, unwanted participation, impedance mismatch, or additional decay channels. 
In nearest-neighbor layouts, loss can accumulate over many hops. 
In hub-based layouts, the number of hops is smaller, but the shared communication layer may dominate the loss budget. 
Hybrid layouts trade distributed link loss against loss in a smaller number of higher-level routing elements. 
Thus graph distance is not a sufficient proxy for error: a short route through a lossy shared element can be less favorable than a longer route through cleaner local links.

\emph{Crowding.}
Increasing the number of modules, ports, couplers, and control lines also increases spectral and geometric crowding. 
More coupled elements raise the probability of frequency collisions, parasitic hybridization, and crosstalk, while denser feedlines and package transitions complicate layout and assembly. 
These constraints are especially important for bosonic hardware, where electromagnetic cleanliness is part of the value of the module. 
A scalable architecture should therefore avoid moving all complexity into a dense central region, even if that centralization improves graph-theoretic distance.

\emph{Calibration.}
A modular processor must remain calibratable as it grows. 
Modularity can help by allowing local calibration procedures to be repeated across nominally identical units, but inter-module couplings introduce nonlocal dependencies. 
Hub-based layouts are most exposed to global calibration correlations because many operations depend on the same shared element. 
Nearest-neighbor layouts distribute these dependencies, but nonlocal operations may require route-dependent tuning. 
Hybrid layouts aim to preserve a mostly local calibration structure while restricting global coordination to a smaller set of gateway modules and higher-level links. 
In this sense, the goal is not simply to minimize the number of calibrated parameters, but to maintain a hybrid calibration problem.

\emph{Robustness.}
Finally, a useful modular processor should tolerate fabrication variation, imperfect interfaces, degraded links, and nonuniform module performance. 
Distributed nearest-neighbor layouts avoid a single dominant communication element, but long routes may be sensitive to any weak link along the path. 
Hub-based layouts simplify routing, but can introduce common-mode bottlenecks or single points of failure. 
Hybrid layouts offer an intermediate failure structure: local operation within a tile can remain available even if some higher-level communication resources are degraded.

These considerations show that graph-theoretic advantage does not translate directly into device-level performance. 
For TBM-based processors, the preferred interconnect is the one that provides the required communication throughput while preserving module coherence, calibratability, and robustness under the intended workload.

\section{Conclusion and outlook}
We have developed an architecture-level framework for modular 3D bosonic quantum processors assembled from repeatable coupled-cavity modules. 
Using graph-theoretic proxies for connectivity, communication distance, routing concentration, and scheduling depth, the analysis shows that interconnect topology does not eliminate the cost of scaling, but redistributes it among local port requirements, multi-hop transport, and shared-resource burden.

The asymptotic comparison and finite-size case studies further show that topology alone does not determine the preferred architecture.
The relevant design depends on physical link quality, available bandwidth and multiplexing, gateway capacity, and the locality of the target workload. 
In particular, architectural advantages predicted by shorter paths or reduced link count must be weighed against loss, crowding, calibration complexity, and contention in the corresponding physical implementation.

A quantitative next step is therefore to connect these graph proxies to device-level models of coupling rates, transfer fidelity, insertion loss, parasitic participation, crosstalk, and communication throughput, together with explicit routing, arbitration, and calibration protocols.
Such a co-design framework will determine which interconnect structures best preserve the coherence and repeatability of the local bosonic modules while supporting useful processor-scale operations.

\begin{acknowledgments}
We thank Yao Lu and Hank Lamm for helpful comments and suggestions that improved this work.
This work was supported by the U.S. Department of Energy, Office of Science, National Quantum Information Science Research Centers, Superconducting Quantum Materials and Systems Center (SQMS), under Contract No. 89243024CSC000002. Fermilab is operated by Fermi Forward Discovery Group, LLC under Contract No. 89243024CSC000002 with the U.S. Department of Energy, Office of Science, Office of High Energy Physics.
\end{acknowledgments}

\appendix

\section{Nearest-neighbor graph metrics}\label{app:nn_metrics}
We consider a \(n\times n\) square array with \(N=n^2\) modules. Its routing graph is the Cartesian product \(G_{\mathrm{NN}}=P_n \square P_n,\) where each module is connected only to its horizontal and vertical nearest neighbors. 
The architecture contains no auxiliary shared routing resource; all routing occurs on the module graph itself.

\subsection{Link count and module degree}
Each of the \(n\) rows contains \(n-1\) horizontal links, and each of the \(n\) columns contains \(n-1\) vertical links.
The total number of communication links is
\begin{equation}
    W_{\mathrm{NN}}=|E_{\mathrm{NN}}|=2n(n-1)=2N-2\sqrt{N}.
\end{equation}
The mean module degree follows from the handshaking lemma:
\begin{equation}
    \overline{K}_{M,\mathrm{NN}}=\frac{2|E_{\mathrm{NN}}|}{N}=4-\frac{4}{\sqrt{N}}.
\end{equation}
For \(n\geq3\), the interior modules have the maximum degree
\begin{equation}
    \Delta_{M,\mathrm{NN}}=4.
\end{equation}
Because the architecture contains no shared auxiliary router or gateway,
\begin{equation}
    \Delta_{I,\mathrm{NN}}=0, \qquad 
    \bar b_{\mathrm{NN}}\equiv\frac{\Delta_{I,\mathrm{NN}}}{N}=0.
\end{equation}

\subsection{Communication distance}
For an open square lattice, the shortest route between modules \(u=(x,y)\) and \(v=(x',y')\) contains \(|x-x'|\) horizontal hops and \(|y-y'|\) vertical hops. 
The shortest-path distance is the Manhattan distance, \(d(u,v)=|x-x'|+|y-y'|,\) where the discrete coordinates \(x,y,x',y'\in\{1,\ldots,n\}\).

We first consider a single one-dimensional coordinate along the \(x\)-axis. 
For a separation \(k=|x-x'|\), there are \(n-k\) unordered coordinate pairs at that separation and \(2(n-k)\) ordered pairs. 
The sum of all ordered one-dimensional distances is
\begin{equation}\label{eq:1D_distance}
    \begin{aligned}
        S_1(n)&=\sum_{x=1}^{n}\sum_{x'=1}^{n}|x-x'|=2\sum_{k=1}^{n-1}k(n-k)\\
              &=2\left[\frac{n^2(n-1)}{2}-\frac{n(n-1)(2n-1)}{6}\right]\\
              &=\frac{n(n^2-1)}{3}.
    \end{aligned}
\end{equation}
For the horizontal contribution \(|x-x'|\), every pair \((x,x')\) occurs for all \(n^2\) choices of \((y,y')\). 
The horizontal contribution to the total distance sum is therefore \(n^2 S_1(n)\). 
By symmetry, the vertical contribution is identical. 
The sum of all ordered two-dimensional distances is
\begin{equation}
    \begin{aligned}
        S_2(n)&=\sum_{u,v}d(u,v)=\sum_{y,y'}\sum_{x,x'}\left(|x-x'|+|y-y'|\right)\\
              &=n^2S_1(n)+n^2S_1(n)=2n^2\frac{n(n^2-1)}{3}.
    \end{aligned}
\end{equation}

The diagonal (self-distance) terms \(u=v\) contribute zero distance. 
Using the convention that the source and destination are uniformly chosen from ordered pairs of distinct modules gives
\begin{equation}
    \begin{aligned}
        \bar{\ell}_{\mathrm{NN}}&=\frac{1}{N(N-1)}\sum_{u\neq v}d(u,v)\\
                                &=\frac{2n^3(n^2-1)}{3n^2(n^2-1)}=\frac{2}{3}\sqrt{N}.
    \end{aligned}
\end{equation}

The largest shortest-path distance occurs between opposite corners, for example between \((1,1)\) and \((n,n)\). 
It is the graph diameter,
\begin{equation}
    \ell_{\mathrm{NN}}^{\text{max}}=|n-1|+|n-1|=2(\sqrt{N}-1).
\end{equation}
Both \(\bar \ell_{\rm NN}\) and \(\ell_{\rm NN}^{\rm max}\) grow as \(\mathcal{O}(\sqrt{N})\), showing that a spatially local lattice shifts the scaling burden into multi-hop nonlocal communication.

In the absence of an auxiliary shared routing resource \(\bar b_{\mathrm{NN}}=0\), the locality is
\begin{equation}
    \eta_{\mathrm{loc,NN}}\equiv1-\bar b_{\mathrm{NN}}=1,
\end{equation}
and the communication coordinate is
\begin{equation}
    \eta_{\mathrm{comm,NN}}\equiv\frac{1}{1+\bar{\ell}_{\mathrm{NN}}}=\frac{3}{3+2\sqrt{N}},
\end{equation}
The illustrative exposure proxy therefore is
\begin{equation}
    \mathcal{E}_{\mathrm{NN}}=1-\eta_{\mathrm{loc,NN}}\eta_{\mathrm{comm,NN}}=\frac{2\sqrt{N}}{3+2\sqrt{N}}.
\end{equation}
These normalized quantities visualize the topology-induced distance--concentration tradeoff; they are not direct physical error probabilities.

\subsection{Path transmission and loss exposure}
To connect graph distance to a simple engineering model, we assume that every nearest-neighbor hop has the same effective transmission efficiency \(0<\eta_0<1\). 
Under the independent-hop approximation, a route of length \(d(u,v)\) has transmission
\begin{equation}
    \eta_{\mathrm{NN}}(u,v)=\eta_0^{d(u,v)}=\eta_0^{|x-x'|}\eta_0^{|y-y'|}.
\end{equation}
This is a scalar transmission model, rather than a complete noise model for a routed quantum operation; correlated noise, nonidentical links, and error correction require a more detailed treatment.

The one-dimensional transmission sum is
\begin{equation}
    S_1(\eta_0)\equiv\sum_{x=1}^{n}\sum_{x'=1}^{n}\eta_0^{|x-x'|}.
\end{equation}
The \(n\) terms with \(x=x'\) contribute unity. 
For each separation \(k\in\{1,\ldots,n-1\}\), there are \(2(n-k)\) ordered coordinate pairs with \(|x-x'|=k\). Therefore,
\begin{equation}\label{eq:1D_transmission_chain_app}
    \begin{aligned}
        S_1(\eta_0)&=n+2\sum_{k=1}^{n-1}(n-k)\eta_0^k\\
                   &=\frac{n(1+\eta_0)}{1-\eta_0}-\frac{2\eta_0(1-\eta_0^n)}{(1-\eta_0)^2}.
    \end{aligned}
\end{equation}
The continuous limit at \(\eta_0=1\) is \(S_1(1)=n^2\).

Because the horizontal and vertical coordinate sums are independent,
\begin{equation}
    \sum_{u,v}\eta_{\mathrm{NN}}(u,v)=\sum_{x,y,x',y'}\eta_0^{|x-x'|}\eta_0^{|y-y'|}=\left[S_1(\eta_0)\right]^2.
\end{equation}
When averaging only over distinct source--destination pairs, the \(N=n^2\) self-pairs have unit transmission and must be removed. 
The exact mean path transmission is
\begin{equation}\label{eq:NN_mean_path_transmission}
    \overline{\eta}_{\mathrm{NN}}=\frac{\left[S_1(\eta_0)\right]^2-N}{N(N-1)}.
\end{equation}
The corresponding mean path-loss probability is
\begin{equation}
    \overline{\mathcal{L}}_{\mathrm{NN}}=1-\overline{\eta}_{\mathrm{NN}}.
\end{equation}
\subsection{Routing concentration}\label{app:nn_betweenness}

For a module \(a\), the shortest-path betweenness is
\begin{equation}
    B_{\mathrm{NN}}(a)=\sum_{\substack{u\neq v\\u,v\neq a}}\frac{\sigma_{uv}(a)}{\sigma_{uv}},
\end{equation}
where \(\sigma_{uv}\) is the number of shortest paths from \(u\) to \(v\), and \(\sigma_{uv}(a)\) is the number of those paths that pass through \(a\) as an intermediate module. 

For a fixed pair \(u\neq v\), there are \(\sigma_{uv}\) shortest paths and \(d(u,v)-1\) intermediate modules after excluding the two endpoints \(u\) and \(v\). 
Summing over all modules and interchanging the order of summation gives
\begin{equation}\label{eq:betweeness_summ}
    \begin{aligned}
        \sum_{a\in V_{\mathrm M}}B_{\mathrm{NN}}(a)
        &=\sum_{u\neq v}\frac{1}{\sigma_{uv}}\left[\sum_{a\in V_{\mathrm M}\setminus\{u,v\}}\sigma_{uv}(a)\right]\\
        &=\sum_{u\neq v}\left[d(u,v)-1\right]\\
        &=N(N-1)(\bar \ell_{\mathrm NN}-1).
    \end{aligned}
\end{equation}
The mean betweenness per module is
\begin{equation}
    \overline{B}_{\rm NN}=\frac{1}{N}\sum_{a\in V_{\mathrm M}}B_{\mathrm{NN}}(a)=(N-1)\left(\bar{\ell}_{\mathrm{NN}}-1\right).
\end{equation}

Because the relative spatial distribution of shortest-path betweenness approaches a size-independent profile in normalized coordinates as the two-dimensional lattice is enlarged, we approximate the asymptotic betweenness distribution by the separable parabolic form
\begin{equation}
    B_{\rm NN}\approx B_{\rm NN}^{\max}\left(1-4x^2\right)\left(1-4y^2\right),\quad
    -\frac{1}{2}\leq x,y\leq\frac{1}{2}.
\end{equation}
The profile is maximal at the center of the lattice and decreases toward its boundary.
Since the normalized square has unit area, the average betweenness is
\begin{equation}
    \overline B_{\rm NN}\approx B_{\rm NN}^{\max}\left[ \int_{-1/2}^{1/2}(1-4x^2)\,dx\right]^2=\frac{4}{9}B_{\rm NN}^{\max}.
\end{equation}

\subsection{Link-conflict scheduling}
The edge chromatic number \(C\equiv\chi'(G)\) is the minimum number of conflict-free activation layers required to activate every physical link once.

For \(n\geq3\), an interior module of \(G_{\mathrm{NN}}\) has four incident edges. 
Since these four edges share a common vertex, they must all receive different colors in any proper edge coloring,
\begin{equation}
    \chi'(G_{\mathrm{NN}})\geq\Delta(G_{\mathrm{NN}})=\Delta_{M,\mathrm{NN}}=4.
\end{equation}
The square lattice is bipartite, so K\"onig's line-coloring theorem guarantees that its edge chromatic number equals its maximum degree
\begin{equation}
    C_{\mathrm{NN}}=\chi'(G_{\mathrm{NN}})=\Delta(G_{\mathrm{NN}})=4.
\end{equation}
The quantity \(C_{\mathrm{NN}}=4\) characterizes the conflict-free scheduling depth for one complete activation of the physical nearest-neighbor links. 

\section{Hub-based architecture}\label{app:hub_metrics}
In this section, we consider a hub-based architecture in which all \(N\) computational modules communicate through one shared routing resource. 
At the graph level, the architecture is the star graph \(G_{\mathrm{Hub}}=K_{1,N},\) with \(N\) module vertices, collected in the set \(V_M\), and one auxiliary hub vertex \(h\). 
The full routing graph has \(N+1\) vertices. 
Each module is a leaf of the star and has no direct module-to-module link.

\subsection{Link count and degree separation}
Since every computational module is connected to the central hub by exactly one link, the total is
\begin{equation}
    W_{\mathrm{hub}}=\left|E_{\mathrm{hub}}\right|=N.
\end{equation}
Each computational module has degree one, whereas the hub has degree \(N\). 
Accordingly, the sum of degrees over the full augmented graph is
\begin{equation}
    \sum_{v\in V_{\mathrm{M}}\cup V_{\mathrm{I}}}\mathrm{deg}(v)=2N=2\left|E_{\mathrm{hub}}\right|,
\end{equation}
consistent with the handshaking lemma.

The module-level degree metrics are evaluated only over the computational modules
\begin{equation}
    \overline{K}_{M,\mathrm{hub}}=\frac{1}{N}\sum_{v\in V_{\mathrm{M}}}\mathrm{deg}(v)=1,
\end{equation}
and
\begin{equation}
    \Delta_{M,\mathrm{hub}}=\max_{v\in V_{\mathrm{M}}}\mathrm{deg}(v)=1.
\end{equation}
Each computational module requires only one communication interface, independent of the total system size.

By contrast, the auxiliary-resource degree is determined by the central hub:
\begin{equation}
    \Delta_{I,\mathrm{hub}}=\max_{h\in V_{\mathrm{I}}}\mathrm{deg}(h)=N.
\end{equation}
The maximum degree of the full routing graph is therefore
\begin{equation}
    \Delta_{\mathrm{hub}}^{\mathrm{max}}=\max\left\{\Delta_{M,\mathrm{hub}},\Delta_{I,\mathrm{hub}}\right\}=N.
\end{equation}

The normalized shared-resource burden is defined as
\begin{equation}
    \bar b_{\mathrm{hub}}\equiv\frac{\Delta_{I,\mathrm{hub}}}{N}=1.
\end{equation}
This quantity measures the largest auxiliary-resource fan-out relative to the total number of computational modules. 

\subsection{Communication distance}
For any two distinct modules \(u,v\in V_M\), the unique shortest path is \(u\rightarrow h\rightarrow v,\) which contains one module-to-hub hop and one hub-to-module hop, \(d(u,v)=2\). 
The mean distance over ordered pairs of distinct modules is therefore
\begin{equation}
    \bar{\ell}_{\mathrm{hub}}=\frac{1}{N(N-1)}\sum_{u\neq v}d(u,v)=2.
\end{equation}

For \(N\geq2\), the largest shortest-path distance is likewise attained between two module leaves \(\ell_{\mathrm{hub}}^{\mathrm{max}}=2\), and both typical and worst-case module-to-module communication distances remain constant as \(N\) grows. 
This hop count represents graph distance only. 

Consequently, the normalized coordinates and the illustrative exposure proxy are
\begin{equation}
    \eta_{\mathrm{comm,hub}}=\frac{1}{3}, \quad
    \eta_{\mathrm{loc,hub}}=0, \quad
    \mathcal{E}_{\mathrm{hub}}=1.
\end{equation}

\subsection{Path transmission and loss}
We assume that every module--hub link has the same effective transmission efficiency \(\eta_0\), and that routing through the hub contributes an additional internal efficiency \(\eta_h\). 
Following the unique shortest path, the corresponding path transmission is
\begin{equation}
    \eta_{\mathrm{hub}}(u,v)=\eta_0\eta_h\eta_0=\eta_0^2\eta_h.
\end{equation}
Because every distinct module pair has the same two-link path, the exact mean path transmission is
\begin{equation}
    \overline{\eta}_{\mathrm{hub}}=\frac{1}{N(N-1)}\sum_{u\neq v}\eta_{\mathrm{path}}(u,v)=\eta_0^2\eta_h.
\end{equation}
The mean path-loss probability is
\begin{equation}
    \overline{\mathcal{L}}_{\mathrm{hub}}=1-\overline{\eta}_{\mathrm{hub}}=1-\eta_0^2\eta_h.
\end{equation}
The worst-case path transmission is identical to the mean value because all module-to-module routes have the same graph distance.

\subsection{Routing Concentration}\label{app:hub_betweenness}
For any two distinct computational modules \(u,v\in V_{\mathrm M}\), there is a unique shortest path between them. 
This path passes through the hub \(h\)
\begin{equation}
    \sigma_{uv}=1, \qquad \sigma_{uv}(h)=1.
\end{equation}
Using the ordered-pair convention adopted here, the betweenness of the hub is
\begin{equation}
    B_{\mathrm{hub}}(h)=\sum_{\substack{u,v\in V_{\mathrm M}\\ u\neq v}}\frac{\sigma_{uv}(h)}{\sigma_{uv}}=N(N-1).
\end{equation}
By contrast, every computational module is a leaf of the star and cannot occur as an intermediate vertex on a shortest path between two other computational modules:
\begin{equation}
    B_{\mathrm{hub}}(u)=0, \qquad u\in V_{\mathrm M}.
\end{equation}
The hub therefore carries the entire module-to-module shortest-path transit burden, whereas the computational modules carry no transit burden. 

\subsection{Link-conflict scheduling}
In the hub graph, no two hub-graph edges can receive the same color in a proper edge coloring. 
Assigning a distinct color to each of the \(N\) module--hub links is a valid proper edge coloring:
\begin{equation}    
    C_{\mathrm{Hub}}=\chi'(G_{\mathrm{Hub}})=\Delta(G_{\mathrm{Hub}})=N.
\end{equation}
This result models the hub as an exclusive routing resource. 

\section{hybrid architecture}\label{app:hybrid_metrics}
We consider a hybrid architecture composed of \(M=N/n\) linear clusters, each containing \(n\) computational modules. 
The \(i\)-th cluster \(\mathcal{C}_i\) contains one designated gateway module \(g_i\), where \(i\in\{1,\ldots,M\}\). 
Each gateway remains part of the computational-module set \(V_{\mathrm M}\), and also serves as the attachment vertex between its local cluster and the higher-level interconnect. 

The higher-level interconnect graph \(H\) contains the gateway terminals and may also contain auxiliary routing vertices \(V(H)=\{g_1,\ldots,g_M\}\cup V_I.\) 
Here, \(V_I\) can represent shared hubs, switches, routers, or other non-module interconnect resources. 
The full hybrid routing graph is obtained by joining the \(M\) local cluster graphs to \(H\) at their gateway vertices:
\begin{equation}\label{eq:hybrid_graph_app}
    G_{\mathrm{hyb}}=\left(V_{\mathrm M}\cup V_{\mathrm I},\cup_{i=1}^M E(\mathcal{C}_i)\cup E(H)\right).
\end{equation}

\subsection{Link count and degree metrics}
We denote by \(|E_H|\) the number of higher-level links, by \(\Delta_{I,\mathrm{hyb}}\) the maximum degree of an auxiliary routing vertex, and by \(b_{\mathrm{hyb}}\) the normalized shared-resource burden. 
These quantities are model-dependent parameters determined by the topology of \(H\). 
For example, in the gateway--hub star a single auxiliary hub \(h\) connects to all \(M\) gateways, we have
\begin{equation}
    |E_H|=M, \quad \Delta_{I,\mathrm{hyb}}=M, \quad \bar b_{\mathrm{hyb}}=M/N.
\end{equation}

To distinguish gateway connectivity from auxiliary-resource connectivity, we define the mean and maximum gateway degrees in \(H\) as
\begin{equation}
    \bar d_{H,g}\equiv\frac{1}{M}\sum_{i=1}^{M}\mathrm{deg}_H(g_i), \quad
    \Delta_{H,g}\equiv\max_{1\leq i\leq M}\mathrm{deg}_H(g_i).
\end{equation}
They count only higher-level links incident on computational gateways, but do not include the degrees of auxiliary routing vertices in \(V_I\). 
In a gateway--hub star, every gateway is connected to the hub by one higher-level link, so that
\begin{equation}
    \mathrm{deg}_H(g_i)=1, \qquad \bar d_{H,g}=1, \qquad \Delta_{H,g}=1.
\end{equation}

The degree of each gateway \(g_i\) in the full hybrid graph is the sum of its local and higher-level degrees
\begin{equation}
    \mathrm{deg}_{G_{\mathrm{hyb}}}(g_i)=\mathrm{deg}_{\mathrm{loc}}(g_i)+\mathrm{deg}_H(g_i),
\end{equation}
where \(\mathrm{deg}_{\mathrm{loc}}(g_i)=2\) for an interior gateway and \(\mathrm{deg}_{\mathrm{loc}}(g_i)=1\) for an endpoint gateway.

Each linear cluster contains \(n-1\) local links, so the total number of local links is
\(|E_{\mathrm{loc}}|=M(n-1)\). 
Since the local and higher-level links are distinct, the total link count is
\begin{equation}
    W_{\mathrm{hyb}}=|E_{\mathrm{loc}}|+|E_H|.
\end{equation}

Each local cluster edge contributes two to the sum of computational-module degrees, whereas each higher-level edge contributes once for each gateway endpoint. 
The total degree of all modules is
\begin{equation}
    \sum_{u\in V_M}\mathrm{deg}_{G_{\mathrm{hyb}}}(u)=2M(n-1)+\sum_{i=1}^{M}\mathrm{deg}_H(g_i).
\end{equation}
Dividing by the number of computational modules gives the exact mean module degree
\begin{equation}
    \overline{K}_{M,\mathrm{hyb}}=\frac{1}{N}\sum_{u\in V_M}d_{G_{\mathrm{hyb}}}(u)=2-\frac{2}{n}+\frac{\bar d_{H,g}}{n}.
\end{equation}

For the gateway--hub star with \(n\ge3\), every linear cluster contains at least one module of local degree two.
The exact maximum computational-module degree is therefore
\begin{equation}
    \Delta_{M,\mathrm{hyb}}=
    \begin{cases}
        3,&\text{interior gateway}\\
        2,&\text{endpoint gateway}
    \end{cases}
\end{equation}

\subsection{Mean communication distance}
In the hybrid architecture described by Eq.~\eqref{eq:hybrid_graph_app}, the mean distance \(\bar \ell_{\mathrm{hyb}}\) is controlled by two contributions: the local distance from a module to its cluster gateway and the higher-level distance between gateways.

Within the same cluster \(\mathcal{C}_i\), the mean distance from a uniformly selected module \(u\) at position \(a\) to a specified gateway \(g_i\) at position \(s\),  including the gateway itself, is
\begin{equation}
    \begin{aligned}
        \bar{\ell}_{g}&\equiv\frac{1}{n}\sum_{u\in\mathcal{C}_i}d(u,g_i)=\frac{1}{n}\sum_{a=1}^n|a-s|\\
                      &=\frac{1}{2n}\left[s(s-1)+(n-s)(n-s+1)\right],
    \end{aligned}
\end{equation}

If the gateway is placed at the center of the cluster, it becomes
\begin{equation}
    \bar{\ell}_{g}=
    \begin{cases}
        \ (n^2-1)/4n, & n \text{ odd},\\
        \ n/4,        & n \text{ even},
    \end{cases}
\end{equation}
 and if the gateway is placed at an endpoint of the cluster, it is
\begin{equation}
    \bar{\ell}_g=(n-1)/2.
\end{equation}
We define the mean distance between distinct gateway terminals as
\begin{equation}
    \bar{\ell}_{H}(M)\equiv\frac{1}{M(M-1)}\sum_{\substack{i,j=1\\i\neq j}}^{M}d_H(g_i,g_j).
\end{equation}

For any fixed source module, there exist $n-1$ potential destinations within the same cluster and $N-n$ destinations across different clusters. 
Assuming all $N-1$ remaining modules are equally likely destinations, the probabilities of intra-cluster and inter-cluster routing are given respectively by
\begin{equation}\label{eq:cluster_probability_app}
    p_{\mathrm{s}}=\frac{n-1}{N-1}, \qquad p_{\mathrm{d}}=\frac{N-n}{N-1}.
\end{equation}

For modules \(u\in\mathcal{C}_i\) and \(v\in\mathcal{C}_j\), with \(i\neq j\), the route has the form
\begin{equation}
    d(u,v)=d(u,g_i)+d_H(g_i,g_j)+d(g_j,v).
\end{equation}
Averaging over the source and destination modules and over distinct cluster pairs gives the mean intercluster distance
\begin{equation}\label{eq:hybrid_mean_distance}
    \bar{\ell}_{\mathrm{hyb}}=p_{\mathrm{s}}\,\bar{\ell}_{\mathrm{loc}}(n)+p_{\mathrm{d}}\left[2\bar{\ell}_{g}(n)+\bar{\ell}_{H}(M)\right],
\end{equation}
where \(\bar{\ell}_{\mathrm{loc}}(n)=(n+1)/3\) is the mean intracluster distance between two distinct modules [see Eq.~\eqref{eq:1D_distance}].

For a gateway--hub star, the \(M\) gateway terminals are connected through one auxiliary hub. Every pair of distinct gateways is separated by two higher-level hops, so \(\bar{\ell}_{H}(M)=2.\) In the balanced case \(M=n=\sqrt{N}\), Eq.~\eqref{eq:hybrid_mean_distance} becomes 
\begin{equation}
    \bar{\ell}_{\mathrm{hyb}}=\frac{1}{3}+\frac{n}{n+1}\left[2\bar{\ell}_{g}(n)+2\right].
\end{equation}

\subsection{Path transmission and loss}
We assume that each local cluster link has an effective power transmission efficiency \(\eta_0\), the gateway-hub link has efficiency \(\eta_H\), and the hub contributes internal routing efficiency \(\eta_h\).

The distance between two  modules \(u\) and \(v\) in the same cluster \(\mathcal{C}_i\) is \(d_{\mathcal{C}_i}(u,v)\), and the transmission within a cluster is \(\eta_{\mathrm{loc}}(u,v)=\eta_0^{d_{\mathcal{C}_i}(u,v)}\).
The sum of the transmissions over all ordered module pairs \(S_1(\eta_0)\) is expressed by Eq.~\eqref{eq:1D_transmission_chain_app}.
Removing the \(n\) diagonal terms with unit transmission, the exact mean intracluster transmission over distinct ordered module pairs is
\begin{equation}
    \overline{\eta}_{\mathrm{loc}}\equiv\frac{S_1-n}{n(n-1)}=\frac{2}{n(n-1)}\sum_{r=1}^{n-1}(n-r)\eta_0^r.
\end{equation}

The intracluster transmission between a uniformly selected module and a gateway is defined as \(\eta_g(u,g_i)=\eta_0^{d_{\mathcal{C}_i}(u,g_i)}\).
Considering the gateway is centrally placed in each cluster, the sum of mean module-to-gateway transmission is
\begin{equation}
    \sum_{x=1}^{n}\eta_0^{|x-g|}=
    \begin{cases}
        1+2\displaystyle\sum_{r=1}^k\eta_0^r, &n=2k+1\\[12pt]
        1+2\displaystyle\sum_{r=1}^{k-1}\eta_0^r+\eta_0^k, &n=2k
    \end{cases}
\end{equation}
and the mean transmission may be written in closed form as
\begin{equation}
    \overline{\eta}_g=
    \begin{cases}
        \frac{1}{n}\left[1+\frac{2\eta_0(1-\eta_0^k)}{1-\eta_0}\right], & n=2k+1,\\[12pt]
        \frac{1}{n}\left[1+\frac{2\eta_0(1-\eta_0^{k-1})}{1-\eta_0}+\eta_0^k\right],& n=2k.
    \end{cases}
\end{equation}

For two modules \(u\in\mathcal{C}_i\) and \(v\in\mathcal{C}_j\) with \(i\neq j\), the route contains the local path from \(u\) to \(g_i\), the two gateway--hub links, one traversal through the hub, and the local path from \(g_j\) to \(v\). 
Its transmission is therefore
\begin{equation}
    \eta_{\mathrm{inter}}(u,v)=\eta_0^{d_{\mathcal{C}_i}(u,g_i)}\eta_H\eta_h\eta_H\eta_0^{d_{\mathcal{C}_j}(v,g_j)}.
\end{equation}
Averaging over all \(n^2\) module pairs associated with a fixed pair of distinct clusters gives
\begin{equation}
    \begin{aligned}
        \overline{\eta}_{\mathrm{inter}}
        &\equiv\frac{1}{n^2}\sum_{u\in\mathcal{C}_i}\sum_{v\in\mathcal{C}_j}\eta_{\mathrm{inter}}(u,v)\\
        &=\eta_H^2\eta_h\Big[\frac{1}{n}\sum_{u\in\mathcal{C}_i}\eta_0^{d_{\mathcal{C}_i}(u,g_i)}\Big]\Big[\frac{1}{n}\sum_{v\in\mathcal{C}_j}\eta_0^{d_{\mathcal{C}_j}(v,g_j)}\Big]\\
        &=\overline{\eta}_g^{\,2}\eta_H^2\eta_h.
    \end{aligned}
\end{equation}
Here, the gateway itself is included among the \(n\) modules in each cluster. 

Average over all distinct ordered module pairs in the full hybrid architecture gives the exact mean path transmission of the gateway--hub hybrid architecture
\begin{equation}\label{eq:hyb_mean_path_transmission}
    \overline{\eta}_{\mathrm{hyb}}
    =p_{\mathrm{s}}\overline{\eta}_{\mathrm{loc}}+p_{\mathrm{d}}\overline{\eta}_{\mathrm{inter}},
\end{equation}
where \(p_{\mathrm{s}}\) and \(p_{\mathrm{d}}\) are defined in Eq.~\eqref{eq:cluster_probability_app}.
Equivalently, the same result may be obtained directly from the total transmission sum. 
Since there are \(M=N/n\) clusters, the total intracluster contribution is \(M\left[S_1-n\right].\)
There are \(M(M-1)\) ordered pairs of distinct clusters, and each ordered cluster pair contributes \(\left(n\overline{\eta}_g\right)^2\eta_H^2\eta_h.\)
Eq.~\eqref{eq:hyb_mean_path_transmission} becomes
\begin{equation}
    \overline{\eta}_{\mathrm{hyb}}=\frac{M\left[S_1-n\right]+M(M-1)n^2\overline{\eta}_g^{\,2}\eta_H^2\eta_h}{N(N-1)}.
\end{equation}

We define the per-link loss probabilities
\begin{equation}
    \epsilon_0\equiv1-\eta_0, \quad \epsilon_H\equiv1-\eta_H, \quad \epsilon_h\equiv1-\eta_h.
\end{equation}
When the accumulated loss along the relevant routes remains small, the first-order expansion gives
\begin{equation}
    \overline{\eta}_{\mathrm{loc}}\simeq 1-\bar{\ell}_{\mathrm{loc}}\epsilon_0,\quad
    \overline{\eta}_g^2\simeq 1-2\bar{\ell}_g\epsilon_0, \quad
    \overline{\eta}^2_H\eta_h\simeq 1-2\epsilon_H-\epsilon_h.
\end{equation}
The mean path-loss probability is therefore
\begin{equation}
    \begin{aligned}
        \overline{\mathcal{L}}_{\mathrm{hyb}}
        &\equiv 1-\overline{\eta}_{\mathrm{hyb}}\\
        &\simeq\epsilon_0\left[p_{\mathrm{s}}\bar{\ell}_{\mathrm{loc}}
        +2p_{\mathrm{d}}\bar{\ell}_g\right]+p_{\mathrm{d}}\left(2\epsilon_H+\epsilon_h\right).
    \end{aligned}
\end{equation}

\subsection{Routing concentration and traffic load}
We consider the balanced row-cluster architecture with a single-hub star realization. 
The architecture contains \(M\) clusters, each consisting of \(n=M\) computational modules, so that \(N=Mn=n^2\).

For two modules \(u\in\mathcal C_i\) and \(v\in\mathcal C_j\) belonging to distinct clusters, there is a unique shortest path between them. 
Every such intercluster path traverses the central hub \(h\), so that
\begin{equation}
    \sigma_{uv}=1,\qquad\sigma_{uv}(h)=1.
\end{equation}
There are \(M(M-1)n^2=N(N-n)\) ordered intercluster module pairs.
The hub betweenness is therefore
\begin{equation}
    B_{\mathrm{hyb}}(h)=N(N-n)=\mathcal O(N^2).
\end{equation}

The gateway modules also carry both intra- and intercluster transit traffic. 
We assume an odd cluster size \(n\), so that the gateway \(g_i\) occupies the center of the one-dimensional chain. 
It divides the remaining modules into two sets of \((n-1)/2\) modules. 
An intracluster shortest path traverses \(g_i\) only when its endpoints lie on opposite sides of the gateway. 
Using ordered source--destination pairs gives
\begin{equation}
    B^{\mathrm{intra}}(g_i)=2\left(\frac{n-1}{2}\right)^2=\frac{(n-1)^2}{2}.
\end{equation}
For intercluster communication, a path traverses \(g_i\) as an intermediate vertex when one endpoint is one of the \(n-1\) nongateway modules in \(\mathcal C_i\) and the other is any of the \(n(M-1)=N-n\) modules outside that cluster. 
Including both ordered directions gives
\begin{equation}
    B^{\mathrm{inter}}(g_i)=2(n-1)(M-1)n=2(n-1)(N-n).
\end{equation}
The gateway betweennes is the sum of both intra- and inter-cluster contributions,
\begin{equation}
    B_{\mathrm{hyb}}(g_i)=\frac{(n-1)^2}{2}+2(n-1)(N-n)=\mathcal O(N^{3/2}).
\end{equation}

\subsection{Link-conflict scheduling}
The scheduling depth \(C_{\rm hyp}\) is determined by the largest module \(\Delta_{M,\mathrm{hyb}}\) or auxiliary-resource \(\Delta_{I,\mathrm{hyb}}\) degree. Since \( \Delta_{M, \mathrm{hyb}}\leq 3\) and \(\Delta_{I, \mathrm{hyb}}=M\) in the gateway-hub star graph, its maximum degree is \(\Delta(G_{\mathrm{hyb}})=M\) for \(M\geq3\). K\"onig's line-coloring theorem gives the exact scheduling depth,
\begin{equation}
    C_{\mathrm{hyb}}=\chi'(G_{\mathrm{hyb}})=M=\frac{N}{n}.
\end{equation}
This scaling has a direct interpretation: all \(M\) gateway--hub links share the same hub and therefore require distinct scheduling layers. 
The bipartite edge coloring guarantees that the local cluster links can be incorporated without increasing the total beyond \(M\) layers.

\section{81-module hierarchical construction}\label{app:hierarchical_81}
The hierarchical architecture consists of nine identical \(3\times 3\) nearest-neighbor tiles arranged in a \(3\times 3\) array. Each tile contains \(n=9\) locally connected modules, including a center gateway module for inter-tile communication, giving \(M=9\) gateways in total. The hybrid metrics in Appendix~\ref{app:hybrid_metrics} are therefore evaluated by replacing the 1D chain sums with the corresponding 2D tile sums.

The number of computational module vertices is
\begin{equation}
    |V_{\rm hier}|=M\times n = 9\times9=81.
\end{equation}
A single \(3\times3\) mesh has \(|E|=12\) links, so the intra-tile links of nine tiles are
\begin{equation}
    |E_{\rm intra}|=9\times 12=108.
\end{equation}
The nine gateway modules are themselves connected as a \(3\times3\) nearest-neighbor mesh at the inter-tile layer and the inter-tile links are \(|E_{\rm inter}|=12\).
The total number of links is therefore
\begin{equation}
    |E_{\rm hier}|=|E_{\rm intra}|+|E_{\rm inter}|=120.
\end{equation}
Thus, the average degree is
\begin{equation}
    \bar K_{\rm hier}=\frac{2|E_{\rm hier}|}{|V_{\rm hier}|}\approx 2.96.
\end{equation}

The maximum degree occurs at the gateway module of the center tile. 
This module has four nearest-neighbor links within its tile and four gateway-layer links to neighboring tiles, giving
\begin{equation}
    \Delta_{\rm hier}^{\max}=4+4=8.
\end{equation}

The probabilities that a destination module lies in the same tile or in a different tile are
\begin{equation}
    p_s=\frac{n-1}{N-1}=0.1, \quad p_d=\frac{N-n}{N-1}=0.9.
\end{equation}
The mean distance between two distinct modules within each tile \(3\times3\) is \(\bar\ell_{\rm loc}=2\), and the mean distance between two distinct gateways on the \(3\times3\) gateway mesh is \(\bar\ell_H=2\). The mean distance from a module to the center gateway is
\begin{equation}
    \bar\ell_g=\frac{1}{9}\sum_{i,j=0}^{2}\left(|i-1|+|j-1|\right)=\frac{4}{3}.
\end{equation} 
Substituting these quantities into Eq.~\eqref{eq:hybrid_mean_distance}, we obtain
\begin{equation}
    \bar\ell_{\rm hier}=0.1\times2+0.9\times(2\times\frac{4}{3}+2)=4.4.
\end{equation}

The maximum local distance to a tile gateway is \(2\), and the maximum distance across the \(3\times3\) gateway mesh is \(4\). The worst-case hierarchical distance is therefore
\begin{equation}
    \ell_{\rm hier}^{\max}=2+4+2=8.
\end{equation}

The mean path transmission in the local \(3\times3\) tile, \(\bar\eta_{\rm loc}\), and in the \(3\times3\) gateway layer, \(\bar\eta_H\), are both calculated from
Eq.~\eqref{eq:NN_mean_path_transmission}. 
In each tile, there is one zero-hop, four one-hop, and four two-hop paths from a module to the center gateway, so the mean module-to-gateway transmission is
\begin{equation}
    \bar\eta_g=\frac{1+4\eta_0+4\eta_0^2}{9}.
\end{equation}
The mean transmission for two modules in different tiles is therefore
\(\bar\eta_{\rm inter}=\bar\eta_g^2\bar\eta_H\).
Using Eq.~\eqref{eq:hyb_mean_path_transmission} with \(\eta_0=\eta_H=0.995\), we obtain
\begin{equation}
    \bar\eta_{\rm path}=p_s\bar\eta_{\rm loc}+p_d\bar\eta_{\rm inter}\approx0.9782 ,
\end{equation}
and the corresponding mean path-loss probability is
\begin{equation}
    \overline{\mathcal{L}}_{\rm hier}=1-\bar\eta_{\rm path}=0.0218
\end{equation}

\bibliography{reference}

\end{document}